\documentclass[prl,aps,twocolumn,groupaddress,nofootinbib,floatfix]{revtex4-1}
\usepackage{latexsym}
\usepackage{graphicx}
\usepackage{verbatim}
\usepackage{multirow}
\usepackage{amsmath}
\newcommand{\beq}{\begin{eqnarray}}
\newcommand{\eeq}{\end{eqnarray}}
\usepackage{mathrsfs}
\usepackage{float}
\usepackage[usenames, dvipsnames]{color}
\usepackage{mathtools}
\usepackage{pgfplots}

\usepackage[ampersand]{easylist}
\usepackage{lhy-shorthand}

\renewcommand \Id {\mathbb{1}}
\newcommand \diag {\operatorname*{diag}}
\newcommand \NN {\nonumber}
\newcommand \up {\uparrow}
\newcommand \dn {\downarrow}

\newcommand \bs {\bar{\sigma}}

\usepackage{slashed}
\usepackage{physics}	% installed packages for typing maths and physics
\usepackage{graphicx}   % need for figures
\usepackage{epstopdf}
\usepackage{subfigure}  % use for side-by-side figures
\usepackage{bbold}
\usepackage{wasysym}

\usepackage[final,unicode,colorlinks=true,citecolor=purple,linkcolor=blue,linktocpage]{hyperref}
\usepackage[obeyFinal,textsize=tiny,backgroundcolor=white,linecolor=magenta]{todonotes}

\begin{document}

\title{Local Entropies across the Mott Transition in an Exactly Solvable Model}

\author{Luke Yeo}
\affiliation{Department of Physics, University of Illinois at Urbana-Champaign, Urbana, Illinois, USA}
\author{Philip W. Phillips}
\affiliation{Department of Physics, University of Illinois at Urbana-Champaign, Urbana, Illinois, USA}

\begin{abstract}
We study entanglement in the Hatsugai-Kohmoto model, which exhibits a continuous interaction-driven Mott transition.
By virtue of the all-to-all nature of its center-of-mass conserving interactions, the model lacks dynamical spectral weight transfer, which is the key to intractability of the Hubbard model for $d>1$.
In order to maintain a non-trivial Mott-like electron propagator, $\SU(2)$ symmetry is preserved in the Hamiltonian, leading to a ground state that is mixed on both sides of the phase transition. % The ground state in this model is a mixed state with both classical and quantum uncertainty. Whereas its quantum uncertainty---in the form of entanglement across bipartitions in position space---indicates itineracy, its classical uncertainty simply originates from an unbroken $\SU(2)$ symmetry.
Because of this mixture, even the metal in this model is unentangled between any pair of sites, unlike free fermions whose ground state carries a filling-dependent site-site entanglement.
We focus on the scaling behavior of the one- and two-site entropies $s_1$ and $s_2$, as well as the entropy density $s$, of the ground state near the Mott transition.
At low temperatures in the two-dimensional Hubbard model, it was observed numerically (Walsh et al., 2018, \href{https://arxiv.org/abs/1807.10409}{arXiv:1807.10409}) that $s_1$ and $s$ increase continuously into the metal, across a first-order Mott transition.
In the Hatsugai-Kohmoto model, $s_1$ acquires the constant value $\ln4$ even at the Mott transition.
The ground state's non-trivial entanglement structure is manifest in $s_2$ and $s$ which decrease into the metal, and thereby act as sharp signals of the Mott transition in any dimension.
Specifically, we find that in one dimension, $s_2$ and $s$ exhibit kinks at the transition while in $d=2$, only $s$ exhibits a kink.
\end{abstract}

\maketitle

\section{Introduction}

It is well known for the Hubbard model that in the vicinity of half-filling, adding and removing electrons changes~\cite{harris,sawatzky,RMP}  the spectrum at all energies.  
This state of affairs obtains because electrons are not the propagating degrees of freedom.  
For example, it has been known since the early work of Harris and Lange~\cite{harris} in 1967 that the low-energy spectral weight is not determined solely by the number of sites, a static quantity, but additionally depends on microscopic parameters in the Hamiltonian, specifically the ratio of the hopping, $t$, to the on-site interaction, $U$.  
This dependence, dubbed dynamical spectral weight transfer (DSWT)~\cite{sawatzky,eskes,RMP}, renders the ground state adiabatically distinct from a Fermi liquid because in such systems no dynamical corrections to the spectral weight exist.  
That is, simply counting electrons exhausts the spectral weight.  
It is this dynamical mixing that makes the Hubbard model non-trivial and gives rise to a slew of non-trivial properties, in particular
(1) an oxygen K-edge absorption~\cite{chen,kedge} spectrum that increases faster than twice the doping level,
(2) an integrated weight of the optical conductivity\cite{uchida,coopernum} in the lower Hubbard band that exceeds the nominal doping level,
and (3) an upper cutoff on the integral of the optical conductivity, for recovery of the superfluid density, of $O(100\Delta)$, $\Delta$, the superconducting gap~\cite{bontemps,marel1,rubhaussen}.  In metals described by Fermi liquid theory, integrating the optical conductivity to $O(\Delta)$ is sufficient to recover the superfluid density.
All such deviations can be understood~\cite{sawatzky,eskes,RMP} within the context of the Hubbard model as a direct consequence of $t/U$ corrections to the spectral weight or the optical conductivity.

As a result of DSWT, exact statements about the $d>1$ Hubbard model are scarce.  
To alleviate this problem, we consider a simplification.  
Such a simplification would be ideal in the context of modern probes of strongly interacting matter such as the entanglement entropy.  
In this paper, we evaluate a measure of the entanglement in an exactly solvable model~\cite{hk1992,baskaran1991exactly} exhibiting a second-order Mott transition.  
In so doing, we show that in addition to the entanglement entropy in free systems, which has been studied extensively~\cite{horodecki2009,huerta,zanardi2002a,zanardi2002,osborne2002,entropy2,entropy3,entropy1,peschel,GK,maldacena}, local entanglement in strongly correlated matter also exhibits key signatures~\cite{tremblay,larsson2006} at phase transitions.  
Whereas the canonical model for such a transition---the Hubbard model---remains intractable in general, the Hatsugai-Kohmoto~\cite{hk1992,baskaran1991exactly} (HK) model is exactly solvable.
The model considers electrons interacting on a lattice with a limited class of all-to-all interactions.
In one space dimension, a scaling analysis shows that both interaction- and density-driven transitions in the HK model lie in the same universality class as the density-driven transition in the Hubbard model~\cite{continentino1992,continentino1994}.
Although markedly different from the Hubbard model, the HK model retains one crucial signature of the Mott transition: a retarded single-particle electron propagator whose real part vanishes at zero energy.
The existence of zeros is the hallmark of Mott insulation~\cite{RMP,zeros1,zeros2}.
Propagators with zeros fail to satisfy the Luttinger sum rule for the ground state~\cite{dave2013} and hence are not adiabatically connected to Fermi liquids.

In this note, we analyze the Mott transition in the HK model from the perspective of local entropies.
These are the two-point entanglement entropy between a pair of lattice sites, the entropy density $s$, and the single-site (two-site) entropy $s_1$ ($s_2$) of the ground state reduced to one (two) lattice site(s).
Our work is motivated in part by a recent analysis of the latter quantities in the Hubbard model~\cite{tremblay}.
Two salient features of the HK model are its (1) mixed (i.e.\ degenerate) ground state and (2) infinite range interactions, to be contrasted with the Hubbard model's (1) pure ground state and (2) local interactions.
For a pure state, entanglement entropy (across a bipartition in position space) measures delocalization of the wavefunction, which appears as itineracy in linear response~\cite{zanardi2002a,zanardi2002}.
Mixed states, on the other hand, carry both classical and quantum correlations, which are generally difficult to distinguish~\cite{modi2012}.
In low-dimensional Hilbert spaces, however, the entanglement of formation is analytically accessible; we utilize the low-dimensional nature of fermionic modes to compute the entanglement entropy between degrees of freedom localized on a pair of lattice sites.
In the larger Hilbert space setting, we study the local entropies $s_1$ and $s_2$ in relation to the entropy density $s$.
Since the former quantities, $s_1$ and $s_2$, result from position space bipartitions, we expect that they carry some of the uncertainty resulting from tracing over delocalized states in the ground state ensemble.
The latter quantity, $s$, cannot encode such quantum correlations in position space, so any discrepancy between the two suggests that entanglement plays a role in the Mott transition in the HK model.  
It is this discrepancy that we analyze here.

\section{Mottness}

\begin{figure}[htb]
    \centering
    \begin{tikzpicture}
        \begin{axis}[
            clip=false,
            hide y axis,
            axis x line=middle,
            xlabel style={at=(current axis.right of origin), anchor=west},
            xlabel=$U$,
            domain=0:0.5,
            xmin=0, xmax=2, xtick={0.01,1}, xticklabels={$0$, $W$},
            ymin=0, ymax=1]

            \node [above] at (axis cs:0.5,0) {metal};
            \node [above] at (axis cs:1.5,0) {insulator};
            \fill [black] (axis cs:1,0) circle(2pt) node[above] {Mott};
        \end{axis}
    \end{tikzpicture}
    \caption{\label{fig:phase-diagram}Phase diagram of the HK model at zero temperature and half filling, as the interaction strength $U$ is tuned from the non-interacting point $U = 0$ across the metal-insulator transition at $U = W$.}
\end{figure}
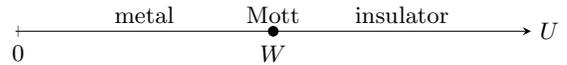

The model~\cite{hk1992,hk1996} we analyze has long-range all-to-all non-local interactions  with standard tight-binding hoppings,
 \begin{eqnarray}\nonumber
	H &=& -t \sum_{\langle j,l\rangle,\sigma} \pqty{ c^\dagger_{j\sigma} c^{}_{l\sigma} + h.c. }  - \mu \sum_{j\sigma} c^\dagger_{j\sigma} c^{}_{j\sigma}\\
&& + \frac{U}{N} \sum_{j_1..j_4}\delta_{j_1+ j_3, j_2+ j_4} c^\dagger_{j_1\uparrow} c^{}_{j_2\uparrow} c^\dagger_{j_3\downarrow} c^{}_{j_4\downarrow},\\ \nonumber
	%=& \sum_{k,\sigma} (\epsilon_k - \mu) c^\dagger_{k,\sigma} c^{}_{k,\sigma} + U \sum_k c^\dagger_{k,\uparrow} c^{}_{k,\uparrow} c^\dagger_{k,\downarrow} c^{}_{k,\downarrow} \\
	%\equiv & \sum_{k,\sigma}\, \xi_k n_{k,\sigma} + U \sum_k n_{k,\uparrow} \, n_{k,\downarrow}
\end{eqnarray}
where the first and second terms denote the local hopping, $t$ and chemical potential, $\mu$. The last term is the infinite-range Hubbard-like interaction $U$; this term is non-zero for electrons that scatter in such a way that their position vectors satisfy the constraint of center of mass conservation given by $j_1+j_3 = j_2 + j_4$.
This model predates the SYK~\cite{sy1993,k2015} model by 2 years,
though it is considerably less studied.
Although both models contain all-to-all non-local interactions, the current model is exactly solvable as a result of the conservation of the center of mass in the interaction term.
The integrability of this model, without resorting to a $1/N$ expansion as in the SYK model~\cite{sy1993,k2015}, is best seen in momentum space
\begin{equation}
	H = \sum_{\vec k}H_{\vec k} = \sum_{\vec k} \left (\xi(\vec k)(\hat{n}_{\vec k\uparrow} + \hat{n}_{\vec k\downarrow} ) + U  \hat{n}_{\vec k\uparrow} \, \hat{n}_{\vec k\downarrow}\right),
\label{eq:kSpaceHK}
\end{equation}
from which it is clear that the kinetic and potential energy terms commute.
In momentum space, the momenta are summed over a square Brillouin zone $[-\pi, \pi)^d$, within which the quasiparticle spectrum $\xi_k = \epsilon_k - \mu$ is set by the dispersion $\epsilon_k = -(W/2d) \sum_{\mu=1}^{d} \cos k^\mu$ with bandwidth $W$ and offset by a chemical potential $\mu$.
Here $n_{k\sigma} = c_{k\sigma}\adj c_{k\sigma}$ is the fermion number operator for the mode with momentum $k$ and spin $\sigma=\up,\dn$.
We consider the system at half filling, fixed by $\mu=U/2$.
As depicted in Fig.~\ref{fig:phase-diagram}, the ground state is metallic for $0<U<W$, insulating for $U>W$, and undergoes an interaction-driven metal-insulator transition at $U=W$.
The phase transition is sharp only at zero temperature, so we work at $T = 0$ throughout.

% For completeness, we also include the single-particle propagator as it contains the occupancies in momentum space which will be used in the entanglement entropy calculations.
The retarded single-particle fermion propagator is related by analytic continuation to the zero temperature Euclidean propagator.
For a fermion in quantum state  $(k,\sigma)$  in the HK model,
\begin{align}\label{eq:propagator}
    G_{k\sigma}(i\omega)
    &\equiv -\int d\tau \ave{c_{k\sigma}(\tau) c_{k\sigma}\adj(0)} e^{i\omega\tau} \\
    &= \frac{1-\ave{n_{k\bar{\sigma}}}}{i\omega - \xi_k} + \frac{\ave{n_{k\bar{\sigma}}}}{i\omega - (\xi_k + U)}
\end{align}
whose pole in the upper (lower) Hubbard band carries a spectral weight equal to the probability $p = \ave{n_{k\bar{\sigma}}}\;$ ($1-p$) that a fermion occupies (does not occupy) the mode with identical momentum $k$ and opposite spin $\bar{\sigma}$.
It is customary to reformulate the Hubbard model~\cite{RMP} in terms of holon $\zeta_{k\sigma} = c_{k\sigma} (1-n_{k\bs})$ and doublon $\eta = c_{k\sigma} n_{k\bs}$ which comprise the fermion $c_{k\sigma} = \zeta_{k\sigma} + \eta_{k\sigma}$.
What distinguishes the HK from the Hubbard model is that the single-particle propagator
\begin{align}
    % G_{k\sigma}^{(\zeta)}(i\omega)
    % &\equiv
    -\int d\tau \ave{\zeta_{k\sigma}(\tau) \zeta_{k\sigma}\adj(0)} e^{i\omega\tau}
    &= \frac{1-\ave{n_{k\bar{\sigma}}}}{i\omega - \xi_k}, \\
    % G_{k\sigma}^{(\eta)}(\tau)
    % &\equiv
    -\int d\tau \ave{\eta_{k\sigma}(\tau) \eta_{k\sigma}\adj(0)} e^{i\omega\tau}
    &= \frac{\ave{n_{k\bar{\sigma}}}}{i\omega - (\xi_k + U)},
\end{align}
is strictly diagonal in terms of these operators because the cross term,
\begin{equation}
    \ave{\zeta_{k\sigma}(\tau) \eta_{k\sigma}\adj}
    = 0
    = \ave{\eta_{k\sigma}(\tau) \zeta_{k\sigma}\adj},
\end{equation}
identically vanishes.
Consequently, the HK model, although it possesses an interaction-driven Mott transition, does not contain DSWT.
As noted previously, it is this feature that makes the model tractable regardless of the spatial dimension.
Whether there are other models that retain this feature but still remain tractable regardless of the spatial dimension is not known at present.
\begin{figure}[tbh]
    \centering
    \begin{tikzpicture}[
    declare function={
        disp(\k) = -(1/2)*cos(\k r);
        invdisp(\E) = rad(acos(-2 * \E));
    }]
    \newcommand\UPLOT{0.7}
    \begin{axis}[
        clip=false,
        axis x line=middle,
        y post scale=0.6,
        hide y axis,
        xlabel=$k$,
        domain=-pi:pi,
        xlabel style={at=(current axis.right of origin), anchor=west},
        xmin=-pi-0.2, xmax=pi+0.2, xtick={-3.142,0,3.142}, xticklabels={$-\pi$, $0$, $\pi$},
        ymin=-0.6, ymax=0.6+\UPLOT, ytick={0}]

        \addplot [domain=-pi:pi] {disp(x)} node[right, pos=1] {$\epsilon_k$};
        \addplot [domain={-invdisp(\UPLOT/2)}:{invdisp(\UPLOT/2)}, line width=2pt]
            {disp(x)};

        \addplot [domain=-pi:pi] {disp(x) + \UPLOT} node[right, pos=1] {$\epsilon_k + U$};
        \addplot [domain={invdisp(\UPLOT/2)-pi}:{pi-invdisp(\UPLOT/2)}, line width=2pt]
            {disp(x) + \UPLOT};

        \addplot [domain=-pi:pi, loosely dashed] {\UPLOT/2} node[right, pos=1] {$\mu=U/2$};

        \draw [dashed] (axis cs:{ invdisp(\UPLOT/2)-pi}, \UPLOT/2) -- ++(axis direction cs:0,-1);
        \draw [dashed] (axis cs:{pi-invdisp(\UPLOT/2)},  \UPLOT/2) -- ++(axis direction cs:0,-1);

        \draw [decorate, decoration={brace,mirror,amplitude=8pt}]
            (axis cs:{invdisp(\UPLOT/2)-pi}, \UPLOT/2-1) -- (axis cs:{pi-invdisp(\UPLOT/2)}, \UPLOT/2-1)
            node[midway, below=\pgfdecorationsegmentamplitude] {$\Omega_2$};
        \draw [decorate, decoration={brace,mirror,amplitude=8pt}]
            (axis cs:{-invdisp(\UPLOT/2)}, \UPLOT/2-1) -- (axis cs:{invdisp(\UPLOT/2)-pi}, \UPLOT/2-1)
            node[midway, below] (left) {};
        \draw [decorate, decoration={brace,mirror,amplitude=8pt}]
            (axis cs:{pi-invdisp(\UPLOT/2)}, \UPLOT/2-1) -- (axis cs:{invdisp(\UPLOT/2)}, \UPLOT/2-1)
            node[midway, below] (right) {};
        \node (label) at (axis cs:0.5,-1.2) {$\Omega_1$};
        \draw [-latex] (label.west) to [bend left=20] (left.south);
        \draw [-latex] (label.east) to [bend right=30] (right.south);
    \end{axis}
\end{tikzpicture}
\caption{\label{fig:bands}Upper and lower Hubbard bands of the one-dimensional HK model in the metallic phase $U < W$.
    Shaded segments indicate occupied momenta.
    $\Omega_2$ labels the doubly-occupied region and $\Omega_1$ the singly-occupied region.}
\end{figure}
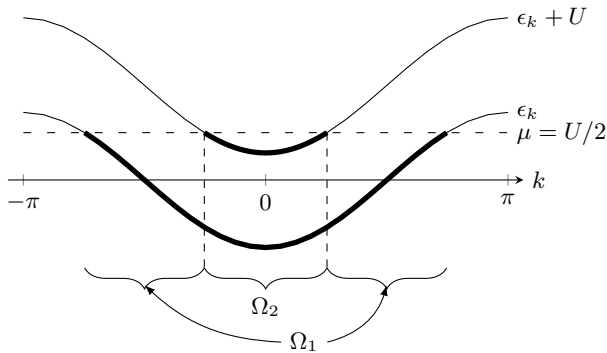

\section{Entanglement}

Consider the ground state produced by the zero temperature limit $\beta \equiv 1/T \to \oo$ of the equilibrium Gibbs state
\begin{equation}
    e^{-\beta H}/Z
    = \bigotimes e^{-\beta H_k} / Z_k,
\end{equation}
where $e^{-\beta H_k} / Z_k$ is the reduced density matrix for the mode $k$.
Here $Z_k = \tr e^{-\beta H_k}$ and
\begin{equation}
    e^{-\beta H_k}
    =
    \begin{pmatrix}
        1 \\
        & e^{-\beta \xi_k} \\
        & & e^{-\beta \xi_k} \\
        & & & e^{-\beta (2\xi_k + U)}
    \end{pmatrix}
\end{equation}
is diagonal in the tensor product basis $\set{\ket{0,0}, \ket{0,1}, \ket{1,0}, \ket{1,1}}$ of the Hilbert space $\cH_{k\up} \ox \cH_{k\dn}$.
Then $e^{-\beta H_k} / Z_k$ is separable across $\cH_{k\up}$ and $\cH_{k\dn}$, and likewise $e^{-\beta H} / Z$ is separable across $\cH_\up = \bigotimes_k \cH_{k\up}$ and $\cH_\dn = \bigotimes_k \cH_{k\dn}$, showing that no entanglement is present between the spin up and down sectors of $e^{-\beta H} / Z$.
Notice however that $e^{-\beta H_k} / Z_k$ cannot be written as $\rho^{k\up} \ox \rho^{k\dn}$, and thereby implements classical correlations between the spin sectors.
We will see that reduced states on one or two sites, instead, have completely uncorrelated spin sectors.
Since $\ave{n_{k\sigma}}$ is a good quantum number, the ground state at zero temperature can be deduced from
\begin{widetext}
    \begin{equation}\label{eq:momentum-mixture}
        \rho^{k}
        \equiv \lim_{\beta \to \oo} e^{-\beta H_k} / Z_k
        =
        \begin{cases}
            \proj{0}_\up \ox \proj{0}_\dn & \text{if } \xi_k > 0 \\
            \half \proj{1}_\up \ox \proj{0}_\dn + \half \proj{0}_\up \ox \proj{1}_\dn & \text{if } \xi_k < 0 \text{ and } \xi_k + U > 0 \\
            \proj{1}_\up \ox \proj{1}_\dn & \text{if } \xi_k < 0 \text{ and } \xi_k + U < 0,
        \end{cases}
    \end{equation}
\end{widetext}
such that modes $k$ with $\xi_k > 0$ are unoccupied, those with $\xi_k < 0$ and $\xi_k + U > 0$ are singly-occupied, and those with $\xi_k < 0$ and $\xi_k + U < 0$ are doubly-occupied.
In the metallic phase with $U<W$, the ground state forms an inner doubly-occupied Fermi volume $\Omega_2$ in which $\ave{n_{k\sigma}} = 1$ and an outer singly-occupied shell $\Omega_1$ in which $\ave{n_{k\sigma}} = 1/2$, as depicted in Fig.~\ref{fig:bands}.
The ground state is indeed half-filled since $2 \abs{\Omega_2} + \abs{\Omega_1} = (2\pi)^d$ is preserved.
In terms of the number of modes $N_i$ in $\Omega_i$, this half-filling condition reads $2 N_2 + N_1 = L^d$, and in terms of the fraction of modes singly- or doubly-occupied $n_i = N_i/L^d$ it reads $2n_2 + n_1 = 1$.
As the phase boundary $U = W$ is approached from the metallic side, $\Omega_2$ vanishes and $\Omega_1$ covers the entire Brillouin zone.
This state persists throughout the insulating phase with $U > W$.
On the other side of the phase diagram, $\Omega_1$ vanishes in the non-interacting limit $U \to 0$.
In each phase, the singly-occupied modes $k \in \Omega_1$ form a mixed sector of the ground state.
The reduced state $\rho^k$ on each singly-occupied mode has non-vanishing mutual information $I({k\up}:{k\dn}) = \ln 2$ between spin sectors.
As a result, $\rho^k$ indeed carries classical correlations while being unentangled.
Distributing the momentum (tensor) product over the mixing sum in $\rho^{k \in \Omega_1}$, we see that the ground state takes the form of a uniform mixture over paramagnetic spin configurations
\begin{align}\label{eq:ground-state}
    \rho
    &= p_\pi
    \sum_{\sum_q \pi(q) = 0}
    \proj{\pi} \\
    \ket{\pi}
    &= \prod_{q \in \Omega_1} c_{q\pi(q)}\adj
    \prod_{k \in \Omega_2} (c_{k\up}\adj c_{k\dn}\adj) \ket{0},
\end{align}
where each of the $\binom{N_1}{N_1/2} = 1/p_\pi$ permutations $\pi: \Omega_1 \to \set{\up,\dn}$ maps the $N_1$ modes in $\Omega_1$ to a paramagnetic spin configuration, resulting in the Bloch state $\ket{\pi}$.
The ground state in the insulating phase has $\Omega_1$ covering the entire Brillouin zone, such that $N_1 = L^d$ and the number of modes $N_2$ in the doubly-occupied region $\Omega_2$ is zero.

Within the singly-occupied region $\Omega_1$, the real part of the retarded propagator at zero temperature takes the form
\begin{equation}
    \Re G_{k\sigma}^{\mathrm{R}}(\omega)
    = \frac{1}{2} \cP \B{ \frac{1}{\omega - \epsilon_k + U/2} + \frac{1}{\omega - \epsilon_k - U/2} },
\end{equation}
given by continuing $i\omega \mto \omega + i0^+$.
At zero energy, the real part $\Re G_{k\sigma}^{\mathrm{R}}(\omega=0)$ vanishes on the surface $\set{k \in [-\pi, \pi)^d: \epsilon_k = 0}$ which always lies inside the region $\Omega_1$.
It is this zero surface~\cite{RMP,zeros1,zeros2} that is the hallmark of Mottness.
Because the region $\Omega_1$ is finite for all couplings $U>0$, the quasiparticle description is valid only at the non-interacting point $U=0$ and breaks down everywhere else in the phase diagram.

From the point of view of position space, each delocalized Bloch wavefunction in the ground state ensemble appears entangled.
Formally the Fourier transform $\bigotimes_k \cH_{k\sigma} \to \bigotimes_j \cH_{j\sigma}$ acts as a global entangling map within each spin sector~\cite{zanardi2002,zanardi2002a}.
Since the ground state $\rho$ is spin-separable, it remains similarly separable after a Fourier transform to position space, and entanglement in $\rho$ can be present only within each spin sector.
Entanglement between the spin-$\sigma$ modes localized on sites $j$ and $j'$ can be determined conclusively from the reduced state $\rho^{jj'\sigma}$ on $\cH_{j\sigma} \ox \cH_{j'\sigma}$.
We refer to these correlations as two-point entanglement.
Following Zanardi's notation in \cite{zanardi2002a}, conservation of particle number mandates that the reduced state be
\begin{equation}\label{eq:two-site-spin-rdm}
    \rho^{jj'\sigma} =
    \begin{pmatrix}
        u \\
        & w_1 & z \\
        & z^* & w_2 \\
        & & & v
    \end{pmatrix}
\end{equation}
in the tensor product basis $\set{\ket{0,0}, \ket{0,1}, \ket{1,0}, \ket{1,1}}$ of $\cH_{j\sigma} \ox \cH_{j'\sigma}$, where the matrix elements are given by
\begin{align}
    z &= \ave{c_{j\sigma}\adj c_{j'\sigma}} \\
    w_1 &= \ave{(1-n_{j\sigma}) n_{j'\sigma}} \\
    w_2 &= \ave{n_{j\sigma} (1-n_{j'\sigma})} \\
    v &= \ave{n_{j\sigma} n_{j'\sigma}} \\
    u &= 1 - w_1 - w_2 - v.
\end{align}
At half filling where $\ave{n_{j\sigma}} = 1/2$, translation invariance obtains
\begin{equation}
    w_1 = w_2 = 1/2 - v
\end{equation}
with $u = v$, and Wick contraction within the ground state ensemble (of pure Bloch wavefunctions) obtains
\begin{equation}
    v = (1/2)^2 - \abs{z}^2.
\end{equation}
That $\abs{z}$ must be sufficiently large, in order that $\rho^{jj'\sigma}$ be entangled, can be seen from the Peres-Horodecki criterion~\cite{horodecki2009}: the two-qubit state $\rho^{AB}$ is separable if and only if its partial transpose $(\rho^{AB})^{\mathsf{PT}}$ has no negative eigenvalues.
Since $\rho^{jj'\sigma}$ is written in the tensor product basis, transposition in the second (inner) Hilbert space $\cH_{j'\sigma}$ can be read off as
\begin{equation}
    \rho^{jj'\sigma} \mto
    (\rho^{jj'\sigma})^{\mathsf{PT}} =
    \begin{pmatrix}
        u & & & z \\
        & w_1 & \\
        & & w_2 \\
        z^* & & & v
    \end{pmatrix}.
\end{equation}
The probability spectrum is mapped to $\set{ w_1, w_2, (1/2)^2 - \abs{z}^2 \pm \abs{z} }$, thereby developing a negative eigenvalue if
\begin{equation}
    \abs{z} > (\sqrt{2}-1)/2 \approx 0.207.
\end{equation}
Turning to momentum space to compute $\abs{z}$, we find that
\begin{align}
    z
    &= \frac{1}{L^d} \sum_{kk'} \ave{c_{k\sigma}\adj c_{k'\sigma}} e^{-i(k\cdot j - k'\cdot j')} \\
    &= \frac{1}{L^d} \sum_{k} \ave{n_{k\sigma}} e^{-ik\cdot(j - j')} \\
    &= \half \frac{1}{L^d} \sum_{k \in \Omega_1} e^{-ik\cdot(j - j')}
    + \frac{1}{L^d} \sum_{k \in \Omega_2} e^{-ik\cdot(j - j')}.
\end{align}
Writing the second term as $\cJ(j-j')$ and the momentum vector with constant components $\vec{\pi}_\mu = \pi$, we reduce the first term to
\begin{align}
    &\quad \frac{1}{L^d} \sum_{k \in \Omega_1} e^{-ik\cdot(j - j')} \NN\\
    &= \frac{1}{L^d} \P{ \sum_{k \in \mathrm{BZ}} - \sum_{k \in \Omega_2} - \sum_{(k-\vec{\pi}) \in \Omega_2} } e^{-ik\cdot(j - j')} \\
    &= 0 - \cJ(j-j') - e^{-i\vec{\pi}\cdot(j-j')} \cJ(j-j') \\
    &=
    \begin{cases}
        0 & \text{if } \norm{j-j'}_1 \text{ is odd}, \\
        -2\cJ(j-j') & \text{if } \norm{j-j'}_1 \text{ is even}.
    \end{cases}
\end{align}
In the latter case, $z$ vanishes.
In the former case, including the adjacent scenario $\norm{j-j'}_1 = 1$, $z = \cJ(j-j')$.

In order to evaluate the domain-restricted sum analytically, we work in the thermodynamic limit near the Mott transition,where the Fermi volumes have spherical symmetry and the sums approach integrals.
As shown in Fig.~\ref{fig:bands}, the boundary of $\Omega_2$ is the locus of $\epsilon_k+U=\epsilon_{\vec{\pi}+k}$.
In the metallic phase near the Mott transition, where $U = W(1-\delta u)$ for small $\delta u > 0$, $\Omega_2$ is a $d$-dimensional ball $B^d(k_{F,2}, \vec{0})$ centered on the origin with radius $k_{F,2} = \sqrt{2d(1-U/W)} \equiv \sqrt{2d \;\delta u}$.
In this regime, the fraction of modes that are doubly-occupied $n_2 = N_2/L^d \ll 1$ is a natural small parameter.
The domain-restricted integral reduces to
\begin{align}
    \cJ(j-j')
    &\to \int_{k \in \Omega_2} \frac{d^dk}{(2\pi)^{d}} e^{-ik\cdot(j - j')} \\
    &= \P{\frac{k_{F,2}/2\pi}{\abs{j-j'}}}^{d/2} J_{d/2}(k_{F,2} \abs{j-j'}) \label{eq:bessel} \\
    &\sim n_2
\end{align}
at leading order in $n_2 \ll 1$.
Then near the Mott transition, $n_2 \not> (\sqrt{2}-1)/2$ so $\rho^{jj'\sigma}$ is separable at this point.
Only in one dimension, where Eq.~(\ref{eq:bessel}) holds also away from the Mott transition, does $\cJ(j-j' = 1)$ cross the threshold value $(\sqrt{2}-1)/2$, deep in the metallic phase at $U/W \approx 0.7835$.
In higher dimensions, Eq.~(\ref{eq:bessel}) provides a good estimate for $\cJ(j-j')$, both of which remain below $(\sqrt{2}-1)/2$ everywhere in the phase diagram.
The ground state therefore remains devoid of two-point entanglement in all dimensions above one and remains separable across the Mott transition in one dimension.

Consider now the local entropies.
As seen above, even the ground state in this model is a mixed state with both classical and quantum uncertainty.
Whereas its quantum uncertainty---in the form of entanglement across bipartitions in position space---indicates itineracy, its classical uncertainty simply originates from an unbroken $\SU(2)$ symmetry.
Now the presence of both classical and quantum uncertainty in the ground state $\rho$ makes it difficult to isolate either portion~\cite{modi2012}, leaving us with hints of itineracy that are muddled by classical uncertainty.
In this context, we consider two local entropies that signal the Mott transition---the entropy density $s$ and the two-site entropy $s_2$---as well as one that does not signal the transition---the single-site entropy $s_1$.
The local entropies,
\begin{align}
    s_1 &\equiv S(\rho^{j}) = -\tr (\rho^j \ln \rho^j), \\
    s_2 &\equiv S(\rho^{\ave{jj'}}) = -\tr (\rho^{\ave{jj'}} \ln \rho^{\ave{jj'}}),
\end{align}
are von Neumann entropies of the reduced states $\rho^{j} = \tr_{i \neq j} \rho$ and $\rho^{\ave{jj'}} = \tr_{i \neq j,j'} \rho$ associated with the bipartitions $\cH_{j} \ox \cH_{\bar{j}}$ and $(\cH_{j} \ox \cH_{j'}) \ox \cH_{\bar{j,j'}}$ across position space, with $j$ and $j'$ neighbouring and the overline denotes the set complement on the lattice, whereas
\begin{equation}
    s \equiv \frac{1}{L^d} S(\rho) = -\frac{1}{L^d} \tr (\rho \ln \rho)
\end{equation}
is the entropy density of the full many-body ground state $\rho$ and does not involve any bipartition of degrees of freedom.
$s_1$ and $s_2$ are not entanglement entropies because the ground state $\rho$ is not pure.%
\footnote{We are unable to perform a conclusive analysis of the entanglement structure of $\rho$, as was done for $\rho^{jj'\sigma}$, because the involved Hilbert space dimensions ($4 \ox 4^{L^d-1}$ and $4^2 \ox 4^{L^d-2}$) are larger than $2 \ox 2$ and $2 \ox 3$.}
$s_2$ and $2s$ measure entropy on the same volume of phase space, each one bounded between zero and $\ln (\dim \cH_j)^2 = \ln 16$, so the two quantities are readily comparable.
However only $s_2$ is sensitive to the details of correlations in the ground state, for instance whether they occur in real or momentum space.
% Each is also independent of the choice of ensemble, say $\set{(p_i, \rho_i)}$ or $\set{(s_l, \sigma_l)}$, in the state $\rho = \sum_i p_i \rho_i = \sum_l s_l \sigma_l$.
% Each ensemble corresponds to a particular preparation of the state $\rho$, and induces an expected entropy $\sum_i p_i S(\rho_i) \neq \sum_l s_l S(\sigma_l)$ that is generically distinct.
% Since one cannot, even in principle, access the microscopic preparation of a thermal state, we study the entropies $S(\rho_1)$ and $S(\rho)$ which include the entropy associated with any block diagonal ensemble~\cite{nielsen2010}.
The entropy density is obtained straightforwardly from the decomposition $\rho = \bigotimes_k \rho^{k}$, with $\rho^{k}$ from Eq.~(\ref{eq:momentum-mixture}), as
\begin{align}
    s
    &= \frac{1}{L^d} \sum_k S(\rho^{k}) \\
    &= \frac{1}{L^d} \sum_{k \in \Omega_1} S(\rho^{k}) \\
    &= n_1 \ln 2 \\
    &= \ln 2 - (\ln 4) n_2
\end{align}
having used additivity of entropy in the first line, vanishing entropy of all pure states $\rho^{k \notin \Omega_1}$ in the second, the definition $n_1 = \abs{\Omega_1}/(2\pi)^d = N_1/L^d$ in the third, and the half filling condition $n_1+2n_2=1$ in the final line.

The computation of $s_1$ and $s_2$ follows simply from the factorization of the associated reduced states $\rho^{j} = \rho^{j\up} \ox \rho^{j\dn}$ and $\rho^{\ave{jj'}} = \rho^{jj'\up} \ox \rho^{jj'\dn}$, which can be seen from the factorization of their matrix elements.
That is, for operators $O_\sigma$ localized on the spin-$\sigma$ sector $\bigotimes_{j \in A} \cH_{j\sigma}$ of $\cH_A$, the matrix elements%
\footnote{Recall that the matrix elements of a reduced state $\rho^A$ on $\cH_A$ can be constructed, given a basis $\ket{a} \ox \ket{b}$ of $\cH_A \ox \cH_B$, from the expectation values
    \begin{equation}
        \ave{ \ket{a}\bra{a'} \ox \Id_B }_\rho
        = \ave{ \ket{a}\bra{a'} }_{\rho^A}
        = \qave{a'}{\rho^A}{a}
    \end{equation}
    of operators $\ket{a}\bra{a'} \ox \Id_B$ localized on $\cH_A$.
}
of the reduced state $\rho^A$ on subregion $A$ factorize as
\begin{equation}
    \ave{O_\up O_\dn}_{\rho^A}
    = \ave{O_\up}_{\rho^A} \ave{O_\dn}_{\rho^A}.
\end{equation}
This factorization is shown in the appendix.
Conservation of particle number leaves $\rho^{j\sigma} = \diag(\ave{1-n_{j\sigma}}, \ave{n_{j\sigma}}) = \Id_2/2$, with the latter equality set by half filling, such that $\rho^{j} = \Id_{4}/4$ is maximally mixed with entropy
\begin{equation}
    s_1 = \ln 4.
\end{equation}
This is consistent with a direct computation of the double-occupancy density $\ave{n_{j\up} n_{j'\dn}} = 1/4$,%
\footnote{Hatsugai and Kohmoto~\cite{hk1992} erroneously find a finite-size correction to $\ave{n_{j\up} n_{j'\dn}}$ from inconsistent asymptotics.
    Their calculation amounts to counting $N_1(N_1-1)$ terms in the sum over $\set{k,q \in \Omega_1: k \neq q}$ in the Fourier transform, but evaluating \unexpanded{$\ave{n_{k\up} n_{q\dn}}$} as $(1/2)^2$ in the sum.
    The latter quantity is instead $(1/2)^2 /(1-1/N_1) = \binom{N_1-1}{(N_1-2)/2}/\binom{N_1}{N_1/2}$, found by counting the number of states $\ket{\pi}$ in the ensemble with $\unexpanded{\ave{n_{k\up}}_\pi} = 1 = \unexpanded{\ave{n_{q\dn}}_\pi}$ for fixed $k,q \in \Omega_1$.}
which holds everywhere in the phase diagram and therefore does not signal the Mott transition.
In the Hubbard model, it is precisely this quantity which changes discontinuously across the Mott transition~\cite{tremblay}.
The matrix elements of $\rho^{jj'\sigma}$ were found around Eq.~(\ref{eq:two-site-spin-rdm}), giving its entropy $S(\rho^{jj'\sigma}) \sim \ln 4 - 4 (n_2)^2$ near the Mott transition, such that
\begin{equation}
    s_2 \sim \ln 16 - 8 (n_2)^2.
\end{equation}
 %, in good agreement with direct numerical evaluation at finite size shown in Fig.~\ref{fig:finite-size}.
Relating $n_2 = \abs{\Omega_2}/(2\pi)^d$ to the volume $\abs{\Omega_2}$ of the $d$-ball with radius $k_{F,2}$ yields the scaling $n_2 \sim c_d (\delta u)^{d/2}$ near the Mott transition, where $c_d = (d/2\pi)^{d/2}/\Gamma(d/2+1)$.
Then
\begin{align}
    s_2 &\sim \ln 16 - 8 (c_d)^2 \;\delta u^d, \\
    2 s &\sim \ln 4 - (\ln 16) c_d \;\delta u^{d/2}
\end{align}
near the Mott transition, whereas $s_1 = \ln4$ everywhere in the phase diagram.

\begin{figure}[tbh]
    \centering
    \begin{tikzpicture}[
        declare function={
            s2d1(\x) = (1/ln(16))*(ln(16) - 8 * (1/(2*pi))/(pi/4) *      (1-x));
            s2d2(\x) = (1/ln(16))*(ln(16) - 8 * (2/(2*pi))^2 *           (1-x)^2);
            s2d3(\x) = (1/ln(16))*(ln(16) - 8 * (3/(2*pi))^3/(9*pi/16) * (1-x)^3);
            sd1(\x)  = (1/ln(16))*(ln(2)  - ln(4) * (1/(2*pi))/(pi/4) *      (1-x)^(1/2));
            sd2(\x)  = (1/ln(16))*(ln(2)  - ln(4) * (2/(2*pi))^2 *           (1-x));
            sd3(\x)  = (1/ln(16))*(ln(2)  - ln(4) * (3/(2*pi))^3/(9*pi/16) * (1-x)^(3/2));
        }]
        \begin{axis}[
            clip=false,
            axis x line=middle,
            axis y line=left,
            xlabel=$U$,
            domain=0:2,
            xlabel style={at=(current axis.right of origin), anchor=west},
            xmin=0, xmax=2, xtick={0,1}, xticklabels={$0$, $W$},
            ymin=0, ymax=1.1, ytick={0,0.5,1}, yticklabels={$0$, $\ln(4)$, $\ln(16)$}]

            \addplot [domain=0.5:1, samples=100] {s2d1(x)} node[left, pos=0] {$d=1$};
            \addplot [domain=0.5:1, samples=100] {s2d2(x)} node[below left = -5pt and 0pt, pos=0] {$d=2$};
            \addplot [domain=0.5:1, samples=100] {s2d3(x)} node[left, pos=0] {$d=3$};
            \addplot [domain=1:2] {1} node[right, pos=1] {$s_2$};

            \addplot [domain=0.5:1, samples=100] {2*sd1(x)} node[left, pos=0] {$d=1$};
            \addplot [domain=0.5:1, samples=100] {2*sd2(x)} node[below left = -4pt and 0pt, pos=0] {$d=2$};
            \addplot [domain=0.5:1, samples=100] {2*sd3(x)} node[above left = -3pt and 0pt, pos=0] {$d=3$};
            \addplot [domain=1:2] {0.5} node[right, pos=1] {$2s$};
        \end{axis}
    \end{tikzpicture}
    \caption{\label{fig:local-entropies}Illustrative ground state local entropies in the HK model at half filling, as the interaction strength $U$ is tuned across the Mott transition at $U = W$.
    $s$ is the entropy density of the full ground state $\rho$, and $s_2$ is the entropy of $\rho$ reduced to two neighbouring sites.
    The curves are exact only for $U \geq W$, with those in $U < W$ given by $U \nearrow W$ asymptotics.}
\end{figure}
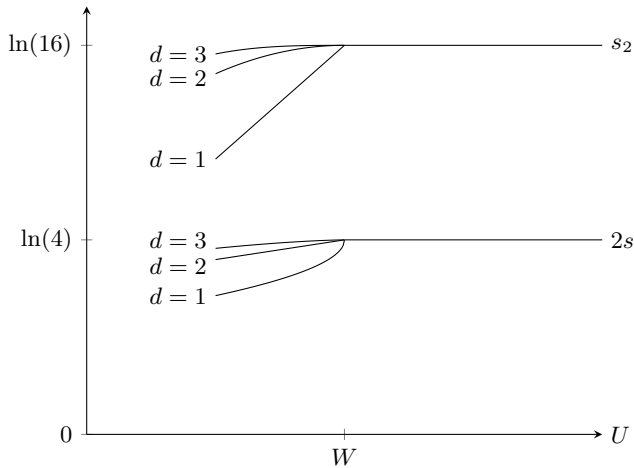

As illustrated in Fig.~\ref{fig:local-entropies}, each local entropy deviates from its insulating value only at $U = W$, thereby signalling the Mott transition in the approach from the insulating phase.
Both local entropies have a kink at the Mott transition in $d=1$, only $s$ has a kink there in $d=2$ dimensions, and both quantities are otherwise smooth there.
The decrease in the entropy density $s$ is entirely explained by a reduced degeneracy of the ground state in the metallic phase.
As discussed earlier, the entropy captured by $s$ is generically mixed into $s_2$, so the decrease in the two-site entropy $s_2$ should, at least in part, be explained likewise.
However, $s_2$ is substantially larger than $2s$ for $U \approx W$ and $U > W$, and their scaling exponents near the Mott transition are different.
The spatial bipartition distinguishing the two local entropies is therefore significant; the decrease in the two-site entropy $s_2$ can be explained independently of the ground state degeneracy.
Given that the single-site entropy $s_1 = \ln4$ is constant, the behaviour of the two-site mutual information
\begin{equation}
    I(j:j')
    \equiv S(\rho^j) + S(\rho^{j'}) - S(\rho^{\ave{jj'}})
    = 2s_1 - s_2
\end{equation}
completely determines the behaviour of the two-site entropy $s_2$, and $I(j:j')$ is itself bounded from below by all connected two-point correlation functions $\ave{O_j O_{j'}} - \ave{O_j}\ave{O_{j'}}$ between sites $j$ and $j'$.%
\footnote{Recall that this bound is sufficiently general to apply also to this mixed state generated by non-local interactions.
    The mutual information $I(A:B)$---between two subsystems $A$ and $B$ in the state $\rho^{AB}$---is a relative entropy $S(\rho^{AB} \unexpanded{\|} \rho^A \ox \rho^B) = I(A:B)$ from the state $\rho^A \ox \rho^B$ with $\rho^{A,B} = \tr_{B,A} \rho^{AB}$, constructed to remove exactly those correlations between $A$ and $B$.
    We use the quantum Pinsker inequality $S(\rho \unexpanded{\|} \sigma) \geq \frac{1}{2} (\norm{\rho - \sigma}_1)^2$ and a H\"older inequality $\norm{\rho^{AB}}_1 \geq \tr (\rho^{AB} O_A O_B)$ for operators $O_I$ supported only on $I = A,B$ and normalized such that its largest singular value $\norm{O_I}_\oo \leq 1$ is bounded by unity.
    Then
    \begin{equation}
        I(A:B) \Bigg|_{\rho^{AB}}
        \geq \frac{1}{2} \P{ \frac{\ave{O_A O_B} - \ave{O_A} \ave{O_B}}{\norm{O_A}_\oo \norm{O_B}_\oo} }
    \end{equation}
    with expectation values taken in the state $\rho^{AB}$.
}
In the insulating phase, $s_2 = 2 s_1$ so the mutual information vanishes and accordingly all connected $j$-$j'$ correlation functions vanish.
From Eq.~(\ref{eq:propagator}) we know that two-point correlations turn on at the Mott transition from gapped insulator to gapless metal, so the mutual information $I(j:j') \gtrsim \ave{c_{j\sigma}\adj c_{j'\sigma}} \sim n_2$ must also turn on there.
Consequently the two-site entropy $s_2$ must decrease from its value in the insulating phase, in a manner governed by two-point correlations.
In the context of correlations and the single-site substructure of $s_2$, the local entropies $s_2$ and $s$ are therefore independent.

\section{Discussion}

Dynamical spectral weight transfer, originating from dynamical double-occupancy, is a fingerprint of the non-trivial propagating degrees of freedom in the Hubbard model.
We have studied the Hatsugai-Kohmoto model of a Mott transition with static double-occupancy, focusing our analysis on two-point entanglement and local entropies in its ground state near the phase transition.
In one dimension, the ground state develops two-point entanglement only deep in the metallic phase, unlike the ground state of free lattice fermions whose two-point entanglement is non-vanishing as long as the lattice is extensively filled~\cite{zanardi2002a}.
In higher dimensions, neither ground state develops two-point entanglement, although the latter is known to possess entanglement between global bipartitions.
Static double-occupancy directly results in a double-occupancy density $\ave{n_{j\up} n_{j\dn}}$ that is constant across the phase diagram, fixing the single-site entropy $s_1$ at the constant value $\ln4$ even at the Mott transition.
On the other hand, the two-site entropy $s_2$ and entropy density $s$ serve as sharp signals of the Mott transition in any dimension.
They are constant in the insulating phase and decrease only when the interaction $U$ is lowered to the transition at $U = W$.
In one dimension, $s_2$ and $s$ feature kinks at the Mott transition, similar to the behaviour of the single-site entropy at phase transitions in the Hubbard chain~\cite{gu2004,larsson2005,larsson2006} and transverse field Ising chain~\cite{osborne2002}; in two dimensions, only $s$ exhibits a kink.
By contrast, it was observed numerically~\cite{tremblay} in the two-dimensional Hubbard model that $s_1$ and $s$ increase continuously as the interaction $U$ is lowered, either into the metal across a first-order Mott transition or in the supercritical crossover region.
This difference arises from the essential component of the Hubbard model, namely the coupling between the low and high-energy operators, namely DSWT.

\acknowledgements

We are thankful to helpful comments from B. Langley and the NSF DMR-1461952 for partial funding of this project.  
\begin{widetext}%

\section{Appendix: Matrix elements of the two-site density matrix}

We compute the matrix elements of the two-site density matrix
\begin{equation}
    \rho^{\ave{jj'}} =
    \mathrm{diag}(
    p_{0,0},
    \rho_{1,0},
    \rho_{0,1},
    \rho_{1,1},
    p_{2,0},
    p_{0,2},
    \rho_{2,1},
    \rho_{1,2},
    p_{2,2}
    )
\end{equation}
block diagonal in the particle number decomposition
\begin{align}
    \cH_{j} \ox \cH_{j'}
    &= \Span \set{ \ket{0,0} } \\
    &\quad\oplus \Span \set{ \ket{\up,0}, \ket{0,\up} }
    \oplus \Span \set{ \ket{\dn,0}, \ket{0,\dn} } \\
    &\quad\oplus \Span \set{ \ket{\up,\dn}, \ket{\dn,\up}, \ket{\up\dn,0}, \ket{0,\up\dn} } \\
    &\quad\oplus \Span \set{ \ket{\up,\up} }
    \oplus \Span \set{ \ket{\dn,\dn} } \\
    &\quad\oplus \Span \set{ \ket{\up\dn,\up}, \ket{\up,\up\dn} }
    \oplus \Span \set{ \ket{\up\dn,\dn}, \ket{\dn,\up\dn} } \\
    &\quad\oplus \Span \set{ \ket{\up\dn,\up\dn} }.
\end{align}

In the one-dimensional subspaces, the diagonal elements are
\begin{align}
    p_{2,2} &= \ave{n_{j\up} n_{j\dn} n_{j'\up} n_{j'\dn}}, \\
    p_{0,0} &= \ave{(1-n_{j\dn}) (1-n_{j'\dn}) (1-n_{j\up}) (1-n_{j'\up})} \\
            &= 2 \ave{n_{j\up} n_{j'\up}} - 4 \ave{n_{j\up} n_{j\dn} n_{j'\up}} + p_{2,2}, \\
    p_{2,0} &= \ave{(1-n_{j\dn}) (1-n_{j'\dn}) n_{j\up} n_{j'\up}} \\
            &= \ave{n_{j\up} n_{j'\up}} - 2 \ave{n_{j\up} n_{j\dn} n_{j'\up}} + p_{2,2}, \\
    p_{0,2} &= p_{2,0}.
\end{align}

In the two-dimensional subspaces,
\begin{align}
    \rho_{2,1}
    &=
    \begin{pmatrix}
        p_{\up\dn,\up} & \zeta \\
        \zeta^* & p_{\up,\up\dn}
    \end{pmatrix} \\
    \text{where } p_{\up,\up\dn} = p_{\up\dn,\up}
    &= \ave{n_{j\up} n_{j\dn} n_{j'\up} (1-n_{j'\dn})} \\
    &= \ave{n_{j\up} n_{j\dn} n_{j'\up}} - p_{2,2}, \\
    \zeta &= \ave{ n_{j\up} n_{j'\up} c_{j\dn}\adj c_{j'\dn}},
\end{align}
and
\begin{align}
    \rho_{1,0} &=
    \begin{pmatrix}
        p_{\up,0} & \zeta' \\
        \zeta'^* & p_{0,\up}
    \end{pmatrix} \\
    \text{where } p_{0,\up} = p_{\up,0}
    &= \ave{(1-n_{j\dn}) (1-n_{j'\dn}) n_{j\up} (1-n_{j'\up})} \\
    &= - \ave{n_{j\up} n_{j'\up}}
    + 3 \ave{n_{j\up} n_{j\dn} n_{j'\up}}
    - p_{2,2}, \\
    \zeta'
    &= \ave{(1-n_{j\dn}) (1-n_{j'\dn}) c_{j\up}\adj c_{j'\up})} \\
    &= \ave{c_{j\up}\adj c_{j'\up}}
    - 2 \ave{ n_{j\dn} c_{j\up}\adj c_{j'\up}}
    + \zeta,
\end{align}
in addition to $\rho_{1,2} = \rho_{2,1}$ and $\rho_{0,1} = \rho_{1,0}$.

In the only four-dimensional subspace,
\begin{equation}
    \quad \rho_{1,1} =
    \begin{pmatrix}
        p_{\up,\dn} & x & w & w^* \\
        x^* & p_{\dn,\up} & -w & -w^* \\
        w^* & -w^* & p_{\up\dn,0} & x' \\
        w & -w & x'^* & p_{0,\up\dn}
    \end{pmatrix}
\end{equation}
where
\begin{align}
    p_{\dn,\up} = p_{\up,\dn}
    &= \ave{ (1-n_{j\dn}) (1-n_{j'\up}) n_{j\up} n_{j'\dn} } \\
    &= 1/4 - 2 \ave{ n_{j\up} n_{j\dn} n_{j'\up} } + p_{2,2} \\
    p_{0,\up\dn} = p_{\up\dn,0}
    &= \ave{ (1-n_{j'\up}) (1-n_{j'\dn}) n_{j\up} n_{j\dn} } \\
    &= 1/4 - 2 \ave{ n_{j\up} n_{j\dn} n_{j'\up} } + p_{2,2} \\
    x &= \ave{ c_{j'\dn}\adj c_{j\up}\adj c_{j\dn} c_{j'\up} } \\
    x' &= \ave{ c_{j\dn}\adj c_{j\up}\adj c_{j'\up} c_{j'\dn} } \\
    w &= \ave{ (1-n_{j'\up}) n_{j\up} c_{j'\dn}\adj c_{j\dn} } \\
      &= \ave{ n_{j'\dn} c_{j\up}\adj c_{j'\up} }^* - \zeta^*
\end{align}

We see that all matrix elements can be written in terms of $p_{2,2}$, $\zeta$, $x$, $x'$, $\ave{c_{j\up}\adj c_{j'\up}}$, $\ave{ n_{j\dn} c_{j\up}\adj c_{j'\up}}$, $\ave{n_{j\up} n_{j'\up}}$, $\ave{n_{j\up} n_{j'\dn}}$, and $\ave{n_{j\up} n_{j\dn} n_{j'\up}}$.
The simplest of these are
\begin{align}
    \ave{n_{j\up} n_{j'\dn}}
    &= \frac{1}{L^{2d}} \sum_{kp}
    p_\pi\sum_\pi \ave{n_{k\up}}_\pi \ave{n_{p\dn}}_\pi \\
    &= p_{\up\dn} \\
    &= 1/4 \\
    &= \ave{n_{j\up}} \ave{n_{j'\dn}}
\end{align}
and $z = \ave{c_{j\up}\adj c_{j'\up}}$, already computed in the text.

Some algebra leads to
\begin{align}
    \ave{n_{j\up} n_{j'\up}}
    &= \frac{1}{L^{2d}} \sum_{k,k'} (1 - e^{-i(k-k')(j-j')})
    \bE_{2}(k,k') \\
    \ave{ c_{j'\dn}\adj c_{j\up}\adj c_{j\dn} c_{j'\up} } = x
    &= -\frac{1}{L^{2d}} \sum_{k,p} e^{-i(k-p)(j-j')}
    \bE_{2}(k,p) \\
    \ave{ c_{j\dn}\adj c_{j\up}\adj c_{j'\up} c_{j'\dn} } = x'
    &= \frac{1}{L^{2d}} \sum_{k,p} e^{-i(k+p)(j-j')}
    \bE_{2}(k,p) \\
    \ave{ n_{j'\dn} c_{j\up}\adj c_{j'\up}} =
    \ave{ n_{j\dn} c_{j\up}\adj c_{j'\up}}
    &= \frac{1}{L^{2d}} \sum_{kp} e^{-ik(j-j')} \bE_{2}(k,p) \\
    \ave{n_{j\up} n_{j\dn} n_{j'\up}}
    &= \frac{1}{L^{3d}} \sum_{k,k',p} (1 - e^{-i(k-k')(j-j')})
    \bE_{3}(k,k',p) \\
    \ave{ n_{j\up} n_{j'\up} c_{j\dn}\adj c_{j'\dn}} = \zeta
    &= \frac{1}{L^{3d}} \sum_{k,k',p}
    e^{-i p (j-j')} (1 - e^{-i (k-k') (j-j')})
    \bE_{3}(k,k',p) \\
    p_{2,2}
    &= \frac{1}{L^{4d}} \sum_{k,k',p,p'}
    ( 1 - e^{-i (k - k') \cdot (j - j')} )
    ( 1 - e^{-i (p - p') \cdot (j - j')} )
    \bE_{4}(k,k',p,p'),
\end{align}
where
\begin{align}
    \bE_2(k,k')
    &= p_\pi \sum_\pi \ave{n_{k\sigma}}_\pi \ave{n_{k'\sigma'}}_\pi \\
    \bE_3(k,k',p)
    &= p_\pi \sum_\pi \ave{n_{k\up}}_\pi \ave{n_{k'\up}}_\pi \ave{n_{p\dn}}_\pi \\
    \bE_4(k,k',p,p')
    &= p_\pi \sum_\pi \ave{n_{k\up}}_\pi \ave{n_{k'\up}}_\pi \ave{n_{p\dn}}_\pi \ave{n_{p'\dn}}_\pi
\end{align}
factorize, for distinct momenta in the thermodynamic limit, as
\begin{align}
    \bE_2(k \in \Omega_a, k' \in \Omega_b)
    &= \frac{a}{2} \frac{b}{2} \\
    \bE_3(k \in \Omega_a, k' \in \Omega_b, p \in \Omega_c)
    &= \frac{a}{2} \frac{b}{2} \frac{c}{2} \\
    \bE_4(k \in \Omega_a, k' \in \Omega_b, p \in \Omega_c, p' \in \Omega_d)
    &= \frac{a}{2} \frac{b}{2} \frac{c}{2} \frac{d}{2}.
\end{align}
Since the phase space of distinct momenta dominates the Fourier sums in the thermodynamic limit, one can read off the above expressions that all matrix elements of the two-site reduced state $\rho^{\ave{jj'}}$ factorize as described in the text, so $\rho^{\ave{jj'}} = \rho^{jj'\up} \ox \rho^{jj'\dn}$ also factorizes.
Then the single-site reduced state also factorizes as $\rho^{j} = \rho^{j\up} \ox \rho^{j\dn}$.

\end{widetext}


%\bibliography{HK-entropy}

\end{document}